\documentclass{elsart}

\usepackage[square,comma]{natbib}
\usepackage{graphicx}
\usepackage{pxfonts}
\usepackage{lineno}

\usepackage{amssymb}

\journal{}

\begin{document}

\thispagestyle{empty}
\begin{Large}
\textbf{DEUTSCHES ELEKTRONEN-SYNCHROTRON}

\textbf{\large{Ein Forschungszentrum der
Helmholtz-Gemeinschaft}\\}
\end{Large}

DESY 10-252

December 2010

\begin{eqnarray}
\nonumber &&\cr \nonumber && \cr \nonumber &&\cr
\end{eqnarray}
\begin{eqnarray}
\nonumber
\end{eqnarray}
\begin{center}
\begin{Large}
\textbf{Circular polarization control for the LCLS baseline in the soft X-ray regime}
\end{Large}
\begin{eqnarray}
\nonumber &&\cr \nonumber && \cr
\end{eqnarray}

\begin{large}
Gianluca Geloni,
\end{large}
\textsl{\\European XFEL GmbH, Hamburg}
\begin{large}

Vitali Kocharyan and Evgeni Saldin
\end{large}
\textsl{\\Deutsches Elektronen-Synchrotron DESY, Hamburg}
\begin{eqnarray}
\nonumber
\end{eqnarray}
\begin{eqnarray}
\nonumber
\end{eqnarray}
ISSN 0418-9833
\begin{eqnarray}
\nonumber
\end{eqnarray}
\begin{large}
\textbf{NOTKESTRASSE 85 - 22607 HAMBURG}
\end{large}
\end{center}
\clearpage
\newpage

\begin{frontmatter}



\title{Circular polarization control for the LCLS baseline in the soft X-ray regime}


\author[XFEL]{Gianluca Geloni\thanksref{corr},}
\thanks[corr]{Corresponding Author. E-mail address: gianluca.geloni@xfel.eu}
\author[DESY]{Vitali Kocharyan}
\author[DESY]{and Evgeni Saldin}

\address[XFEL]{European XFEL GmbH, Hamburg, Germany}
\address[DESY]{Deutsches Elektronen-Synchrotron (DESY), Hamburg,
Germany}

\begin{abstract}
The LCLS baseline includes a planar undulator system, which produces intense linearly polarized light in the wavelength range $0.15$-$1.5$ nm. In the soft X-ray wavelength region polarization control from linear to circular is highly desirable for studying ultrafast magnetic phenomena and material science issues.  Several schemes using helical undulators have been discussed in the context of the LCLS. One consists in replacing three of the last planar undulator segments by helical (APPLE III) ones. A second proposal, the 2nd harmonic helical afterburner, is based on the use of short, crossed undulators tuned to the second harmonic. This last scheme is expected to be the better one. Its advantages are a high (over $90 \%$) and stable degree of circular polarization and a low cost. Its disadvantage is a small output power ($1 \%$ of the power at the fundamental harmonic) and a narrow wavelength range. We propose a novel method to generate $10$ GW level power at the fundamental harmonic with $99 \%$ degree of circular polarization from the LCLS baseline. Its merits are low cost, simplicity and easy implementation. In the option presented here, the microbunching of the planar undulator is used too. After the baseline undulator, the electron beam is sent through a $40$ m long straight section,  and subsequently passes through a short helical (APPLE II) radiator. In this case the microbunch structure is easily preserved, and intense coherent radiation is emitted in the helical radiator. The background radiation from the baseline undulator can be easily suppressed by letting radiation and electron beam through horizontal and vertical slits upstream the helical radiator, where the radiation spot size is about ten times larger than the electron bunch transverse size. Using thin Beryllium foils for the slits the divergence of the electron beam halo will increase by Coulomb scattering, but the beam will propagate through the setup without electron losses. The applicability of our method is not restricted to the LCLS baseline setup. Other facilities e. g.  LCLS II or the European XFEL may benefit from this work as well, due to availability of sufficiently long free space at the end of undulator tunnel.
\end{abstract}

%
%
%
\end{frontmatter}



\section{\label{sec:intro} Introduction}

\begin{figure}
\includegraphics[width=1.0\textwidth]{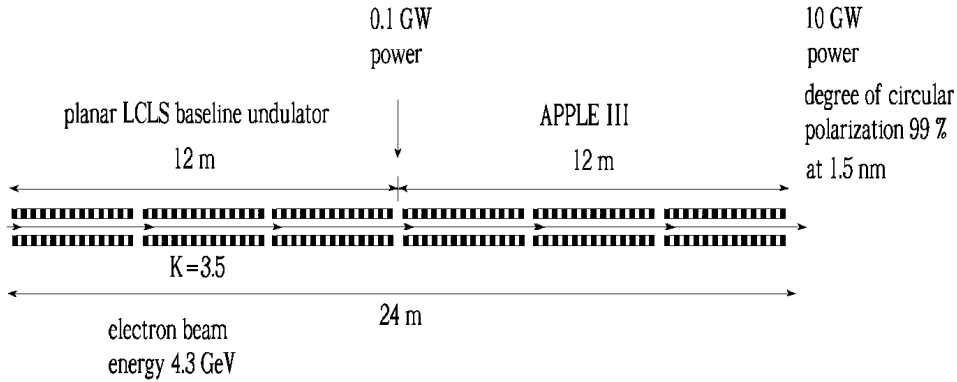}
\caption{The first option for circular polarization control at the LCLS. For the soft X-ray wavelength range, helical polarization can be most effectively achieved by letting the electron bunch pass through a helical undulator. It is not necessary that all the undulators in the line be helical. Once the SASE process has induced microbunching in the electron beam, the microbunches radiates coherently in the helical undulator tuned
at the same wavelength. In order to reach a circular polarization degree larger than $99 \%$  APPLE undulators need to be installed in the linear regime, before the power emitted by the bunching undulator reaches $0.1$ GW.} \label{lclspc7}
\end{figure}

\begin{figure}
\includegraphics[width=1.0\textwidth]{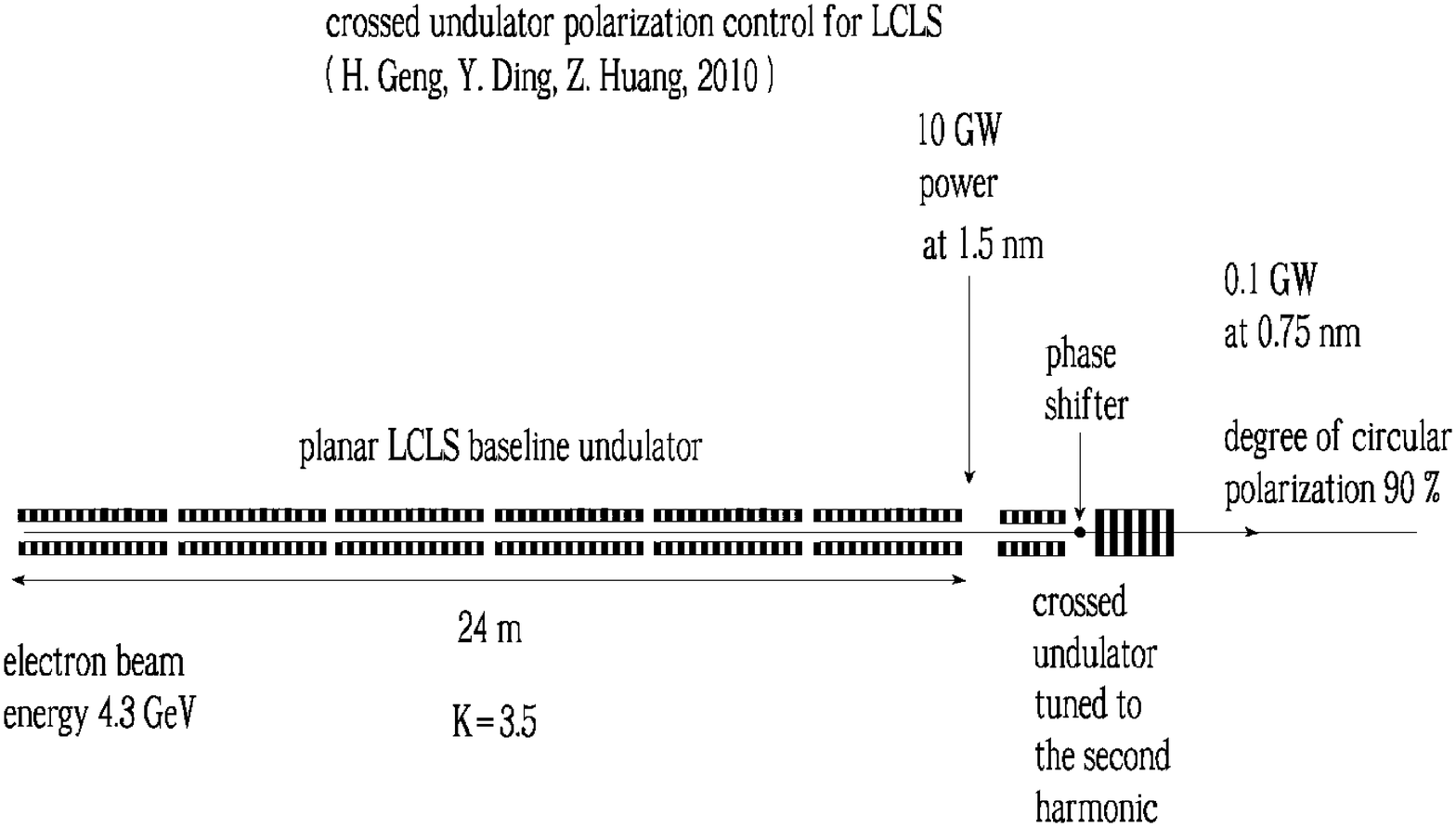}
\caption{The second option for circular polarization control option at the LCLS. The radiation in the baseline undulator is characterized by a different frequency compared to the radiation in the crossed undulator, and thus has no effect on the polarization properties of the harmonic fields. The maximum circular degree of polarization achievable is over $90 \%$ in the SASE regime, and is insensitive to the length of the baseline undulator.} \label{lclspc8}
\end{figure}
Circularly polarized X-ray radiation is a useful tool for investigating magnetic materials and other material science issues, as discussed in \cite{LCLS1}-\cite{tdr-2006} and references therein. However, the LCLS baseline \cite{LCLS2} is composed by a planar undulator system, which produces intense linearly polarized light in the wavelength range $0.15-1.5$ nm. For hard X-ray radiation (shorter than $0.6$ nm), longitudinally coherent, linearly polarized X-rays can be obtained with the help of self-seeding techniques, and converted \cite{OURY4} into any other elliptic polarization state by passing them through X-ray phase retarders \cite{HIRA,FREE}. For soft X-ray wavelengths longer than $0.6$ nm, two main methods have been proposed to achieve circular polarization. The first consists in letting the electron beam through helical undulator devices.  Such devices \cite{BAHR}  are mechanically more complex than the simple, fixed gap planar undulator currently employed at the LCLS, Fig. \ref{lclspc7}. A second possibility, building on the work \cite{KIMI}, is constituted by the use of crossed planar undulators tuned at a different frequency than the fundamental, Fig. \ref{lclspc8}. These options are discussed e.g. in \cite{GENG}.

Let us briefly describe the first option. The option of a full-length helical undulator remains the most attractive in terms of quality of the output radiation. However, its realization is not foreseen at the first stage of the LCLS project mainly due to technical challenges related to the production of long helical insertion devices. The choice of a relatively short helical undulator could initially constitute a reasonable compromise.  It is not necessary that all the undulators in the line be helical, Fig. \ref{lclspc7}. In fact, since the SASE process already provides electron beam microbunching, the microbunches radiates coherently when passing through an helical undulator tuned at the same radiation wavelength. At the LCLS, saturation of linearly polarized radiation at $1.5$ nm  is reached after $6$ undulator modules. The saturation power is  about $10$ GW. In order to reach more than $99 \%$ degree of polarization, APPLE undulators need to be installed before the power reaches $0.1$ GW, i.e. in the fourth module \cite{KUSK}.

Considering the second option, a short pair of crossed planar undulator is placed behind the long planar undulator, Fig. \ref{lclspc8}. The radiation in the baseline undulator is characterized by a different frequency compared to the radiation in the crossed undulator, and thus has no effect on the polarization properties of the harmonic fields. The maximum circular degree of polarization achievable is over $90 \%$ in the SASE regime, and is insensitive to the length of the baseline undulator. A scheme was also presented in \cite{GENG}, where the crossed undulators are tuned to the fundamental. In this case, if the crossed undulator length is $1.3$ times the FEL gain length, the two orthogonal linear components turn out to produce radiation with the same intensities and their combination results in circular polarization if their phase difference is equal to $\pi/2$. The FEL gain length, however, depends on a number parameters such as wavelength, peak current, emittance, energy spread, beta function. Some of them might fluctuate leading to fluctuations of the degree of polarization as well.

In a coaxial setup where the baseline LCLS undulator is followed by a short radiator tuned at the same wavelength, there is the issue of separation of linearly from circularly polarized radiation. In the option presented in \cite{YLI}, after the planar undulator the electron beam is deflected by a bending system and subsequently passed through a helical undulator. The bending system serves to separate the linearly from the circularly polarized radiation. The polarization properties of the radiator in this scheme are completely independent of the light produced in the undulator providing the bunching of the electron beam. Using this scheme, the electron beam microbunching on the scale of the soft X-ray radiation wavelength produced in the planar undulator must be maintained through the bending system, and this constitutes a challenge, which started to be addressed in literature only very recently. According to \cite{YLI}, such a scheme would be as long as 80 m.

\begin{figure}
\includegraphics[width=1.0\textwidth]{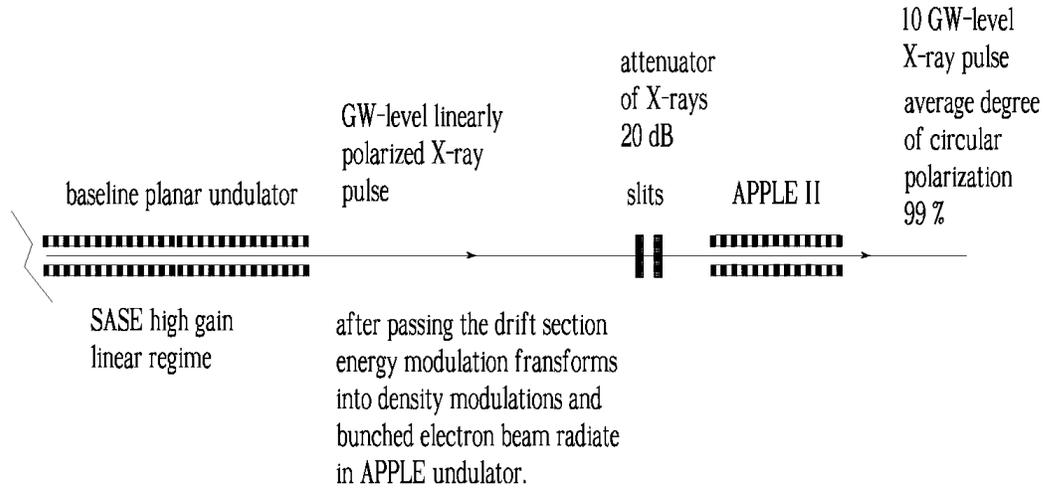}
\caption{Concept of circular polarization control at LCLS baseline. After the baseline undulator the electron beam is propagated along the 40 m
long straight section and subsequently passed through a helical radiator. In this case the microbunch structure of the buncher is preserved and intense coherent radiation is emitted in the helical undulator.  Linearly-polarized radiation  from the baseline undulator is easily suppressed
by spatially filtering out the $99 \%$ of incident power by slits without spoiling of electron beam} \label{lclspc5}
\end{figure}
As a result only two main options were selected as candidates for the LCLS upgrade, Fig. \ref{lclspc7} and Fig. \ref{lclspc8}. In this paper we propose a third option, Fig. \ref{lclspc5}, which mainly consists sending the electron beam, after the passage in the baseline undulator, through a $40$ m long straight section,  and subsequently through a short helical (APPLE II) radiator.  The background radiation from the baseline undulator is suppressed by letting radiation and electron beam through horizontal and vertical slits upstream the helical radiator, where the radiation spot size is about ten times larger than the electron bunch transverse size. Using thin Beryllium foils for the slits the divergence of the electron beam halo will be spoiled due to Coulomb scattering, but the beam will propagate through the setup without electron losses.

Our method presents obvious advantages compared with the previously proposed ones. In all three cases one would obtain very high and stable  degree of polarization, because all three options propose radical solutions of background problem. The option proposed in this paper, however, has advantages over the first one in terms of costs and time, not only because our helical undulator would be shorter, but also because we can afford to use the existing design of APPLE II type undulators, improved for PETRA III \cite{BAHR}, instead of APPLE III type undulators \cite{KUSK}, which have not yet come into operation. Comparing our technique with the crossed-undulator proposal, we have the advantage in terms of power and wavelength range. Finally, to be specific here we restrict ourselves to the investigation of APPLE modules following the main undulator. However, our scheme can also operate with any helical radiator, including e.g. crossed planar undulators allowing for fast helicity switching.

\section{Possible circular polarization control scheme with spatially
filtering out the linearly-polarized radiation from the LCLS baseline
undulator}

\begin{figure}
\includegraphics[width=1.0\textwidth]{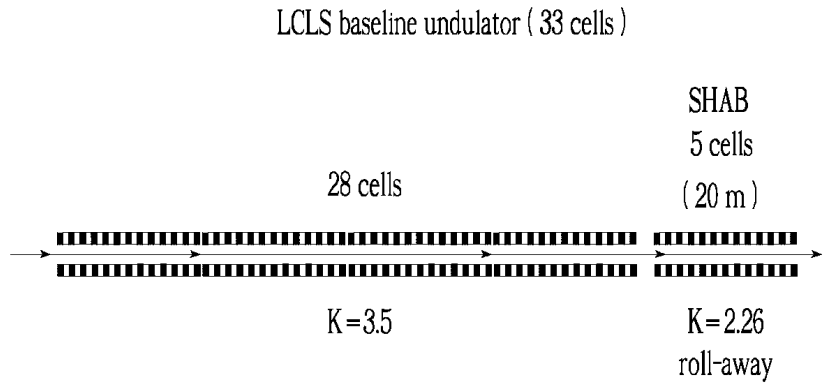}
\caption{Current design of the LCLS baseline undulator system. The support and motion system allows for an undulator to be retracted remotely, leaving the vacuum chamber in place, by 80 mm, and then to be returned to the original position with an accuracy of $2~\mu$m. We assume that last five (SHAB) undulator  modules are rolled away from the beamline in this fashion.} \label{lclspc1}
\end{figure}

\begin{figure}
\includegraphics[width=1.0\textwidth]{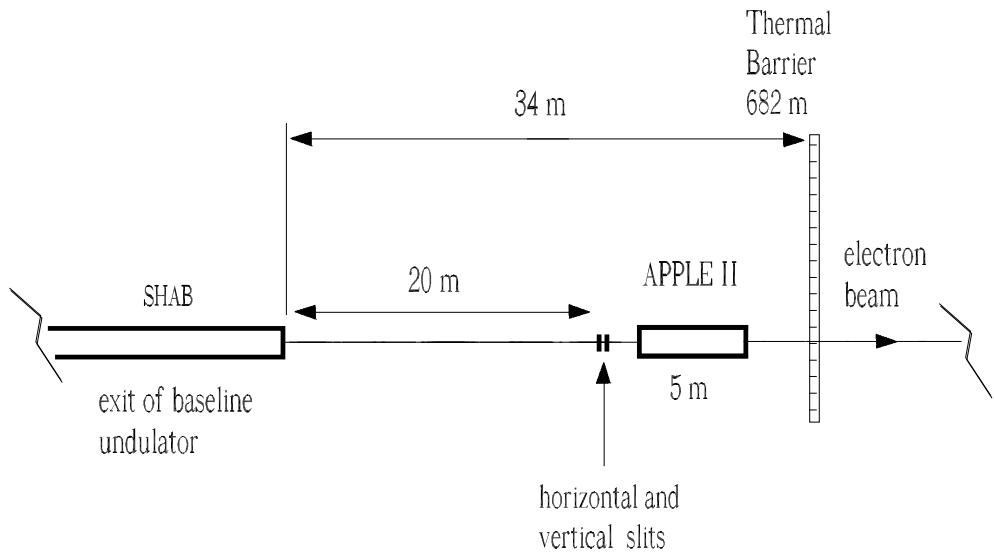}
\caption{The installation of horizontal and vertical slits and of the $5$ m-long APPLE II undulator after the LCLS baseline undulator will allow to produce high-power, highly circularly-polarized soft-X-ray radiation.} \label{lclspc2}
\end{figure}

\begin{figure}
\includegraphics[width=1.0\textwidth]{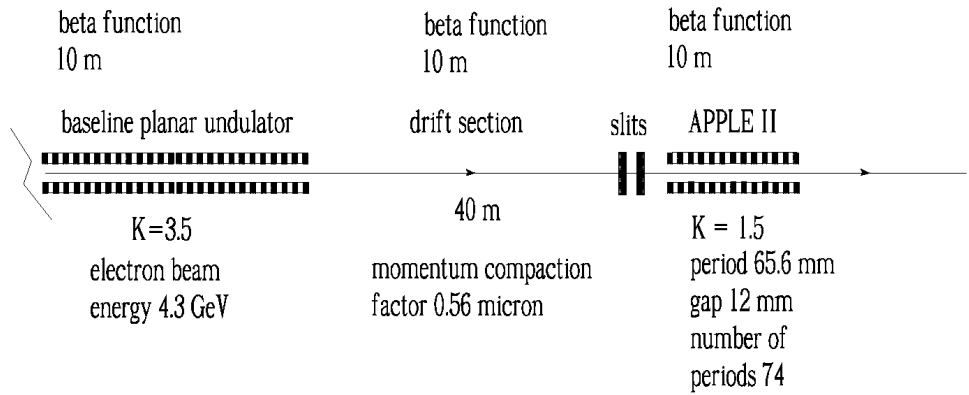}
\caption{Design of the undulator system for circular polarization control at the LCLS
baseline} \label{lclspc4}
\end{figure}

\begin{figure}
\includegraphics[width=1.0\textwidth]{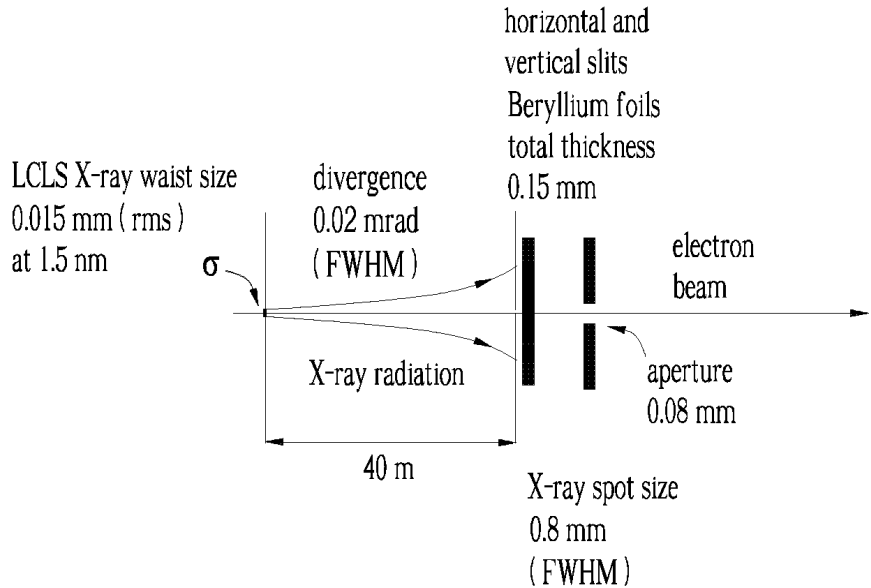}
\caption{Simple method for suppressing linearly-polarized soft X-ray radiation from the LCLS baseline undulator. It is possible to eliminate the linearly polarized background by using a spatial window positioned after the planar undulator exit. This can be practically implemented by
letting radiation and electron bunch through slits at positioned 40 m downstream of the planar undulator, where the radiation pulse has a ten times larger spot size compared with the electron bunch transverse size} \label{lclspc3}
\end{figure}

\begin{figure}
\includegraphics[width=1.0\textwidth]{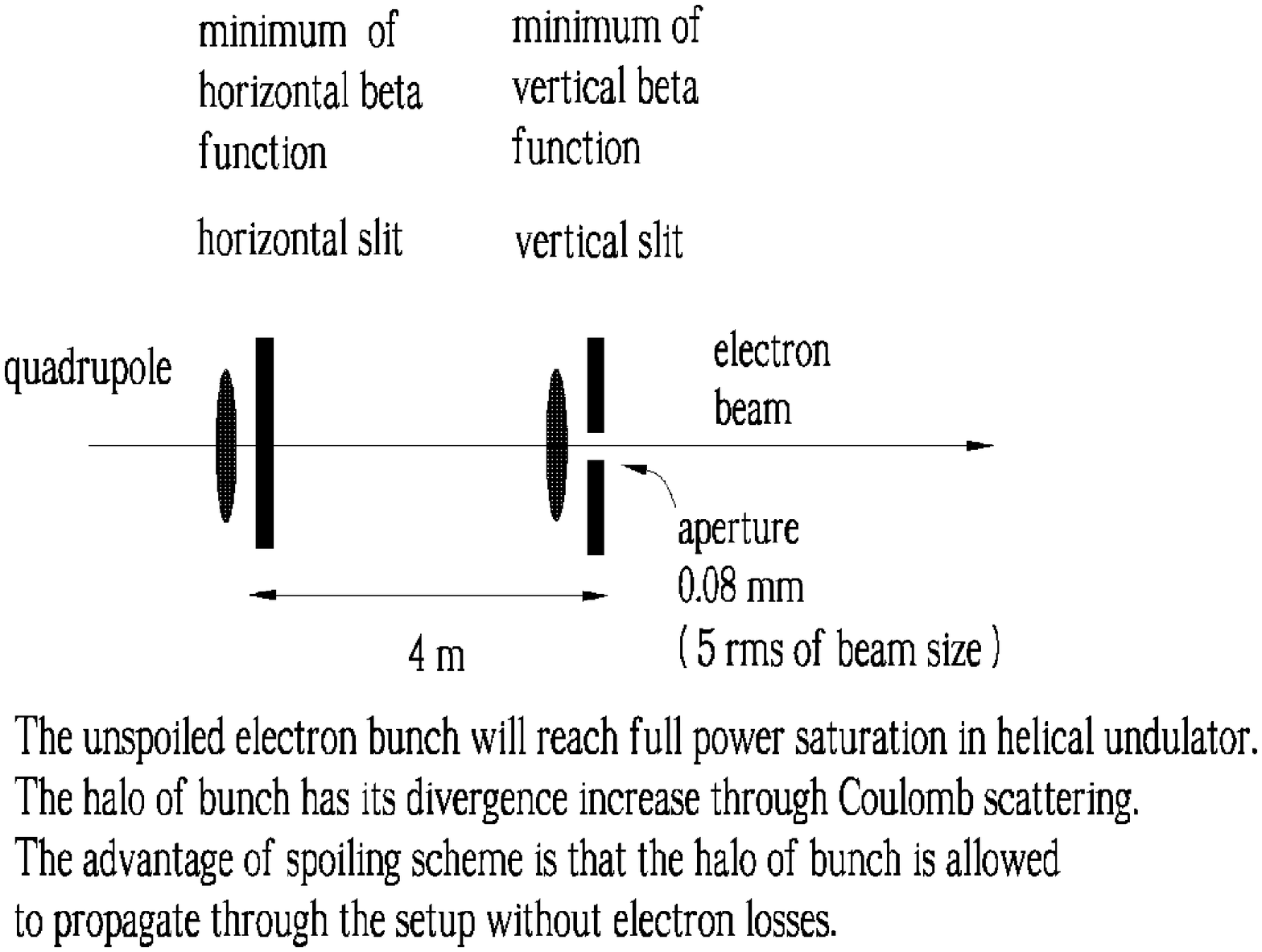}
\caption{Horizontal and vertical slits  (thin Beryllium foils)  as (20 dB) attenuator for linearly polarized radiation from baseline undulator. Only the halo of the electron bunch will be spoiled i.e. will have its divergence increased through Coulomb scattering. The advantage of the spoiling scheme is that the halo of bunch is allowed to propagate through the setup up to the beam dump without electron losses} \label{lclspc6}
\end{figure}

\begin{figure}
\includegraphics[width=1.0\textwidth]{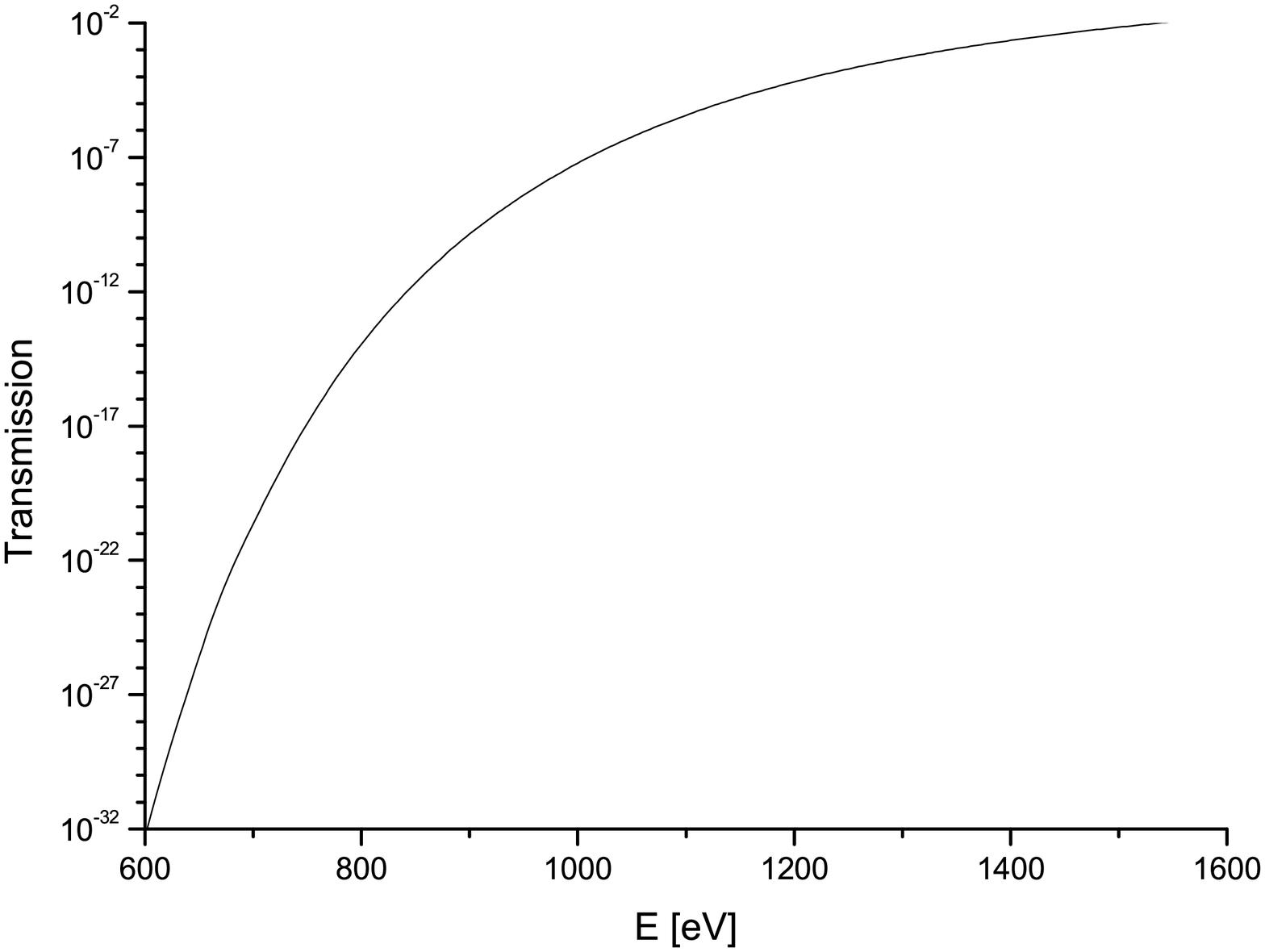}
\caption{Transmission of the a $150~\mu$m-thick Beryllium foil as a function of the energy in the $600$ eV-$1600$ eV energy-range. This thickness is enough to block the radiation from the first LCLS undulator.} \label{transmB}
\end{figure}
The electron beam first goes through the baseline undulator, producing SASE radiation, and inducing energy and density modulation on the electron beam.  This first step is illustrated in Fig. \ref{lclspc1}, where we also assume that the five second harmonic afterburner (SHAB) modules are rolled away \cite{BAIL} from the beamline, Fig. \ref{lclspc2}. In this way, we provide a total of $40$ m straight section for the electron beam, $20$ m corresponding to the SHAB modules, and further $20$ m corresponding to the straight section after the exit of the main undulator. At the end of the $40$ m-long straight section we install horizontal and vertical slits, and a $5$ m-long APPLE II type undulator Fig. \ref{lclspc4}. While passing through this last section, the microbunched electron beam produces intense bursts of radiation in any selected polarization state. However, one should account for the fact that the straight section acts as a dispersive element. Therefore, in a klystron-like bunching effect should also be accounted for, which modifies the density modulation at the exit of the first undulator. From this viewpoint, the first (baseline) LCLS undulator behaves as an energy modulator, and the drift section, i.e. the straight section, transforms the energy into density modulation. Following these lines, before discussing the result of numerical simulations, we present here a qualitative treatment of the influence of the propagation of electron beam through the drift section on the electron beam microbunching.

The way the electron bunch is modulated in an optical klystron is quantitatively described in e.g. \cite{CZON}. The current $I$ at the exit of the straight section is found to be a composition of harmonics of the fundamental frequency $\omega$ according to

\begin{eqnarray}
I &=& I_0 + 2 I_0 \sum_{n=1}^{\infty} \exp\left[-\frac{n^2}{2}
\sigma_E^2 \left(\frac{\omega R_{56}}{c
{E}_0}\right)^2\right]  J_n\left(n P_0 \frac{\omega
R_{56}}{c {E}_0}\right)\cr && \times \cos\left[n \omega
\left(\frac{z}{v_z}-t\right)\right]~, \label{czonca}
\end{eqnarray}
where $P_0$ is the energy modulation after the first undulator, ${E}_0$ is the nominal electron energy, $\sigma_E$ is the local energy spread of electrons, $z$ is the longitudinal coordinate, $v_z$ is the longitudinal velocity of electrons and $t$ is the time. Moreover $J_n$ indicates the Bessel function of the first kind of order $n$ and, as before, $R_{56}$ is the momentum compaction factor. Let us consider the first harmonic bunching in Eq. (\ref{czonca}), $n=1$. We assume  $E_0 = 4.3$ GeV, a fundamental wavelength $\lambda = 1.5$ nm, a straight section length $L=40$ m and $\sigma_E = 1.5$ MeV energy spread. We can estimate $R_{56} \simeq L/\gamma^2 \simeq 560$ nm, with a relative rms energy spread of the $0.03 \%$. The exponential factor in Eq. (\ref{czonca}) turns out to be about $\exp(-0.33) \sim 0.7$, i.e. reasonably near unity. This means that modulation is not destroyed by the passage through the straight section. Moreover, it should be noted that this factor is energy dependent. In particular, increasing the energy will scale down the wavelength as $\sim \gamma^{-2}$,  $R_{56} \sim \gamma^{-2}$ and $\sigma_E/E_0 \sim \gamma^{-1}$. Therefore, the exponential factor scales as $\gamma^{-1}$, so that for wavelength shorter than $1.5$ nm, which is the case under study in our numerical example, the suppression factor will be even smaller.

With the help of Eq. (\ref{czonca}) we can estimate the maximal FEL-induced energy modulation which can be tolerated in the electron beam after the baseline undulator. This can be done requiring that the argument of the Bessel function in Eq. (\ref{czonca}) be not larger than unity. This fixes the maximal value of $P_0 \sim 0.04\%$. At saturation, the energy modulation is comparable with the energy losses of the bunch, which is of the order of the relative bandwidth of the spectrum. In particular, for $\lambda = 1.5$ nm we have a bandwidth, and therefore, energy losses and energy modulation, in the order of $0.5 \%$. Comparing with the maximal value of $P_0 \sim 0.04 \%$, we conclude that the planar undulator should work in the linear regime, and that the FEL power yielded in the first part of our setup should be about ten times smaller than at saturation. Numerical simulations (see Section \ref{sims}), confirm these estimations.

The influence of the betatron motion should be further accounted for. In fact, the finite angular divergence of the electron beam, linked with the betatron function, leads to an additional spread in the longitudinal velocity and as a consequence to an additional suppression factor in Eq. (\ref{czonca}). We can estimate this factor by comparison with the influence of the energy spread. The deviation in longitudinal velocity due to angular deviation is simply found as $\Delta v_z  = v[\cos(\theta) - 1] \sim - v\theta^2/2$. Considering $\theta$ as the rms angular divergence we estimate $\Delta v_z/v \sim - \varepsilon/(2 \beta)  \sim -2.4 \cdot 10^{-12}$, where $\epsilon =4.8 \cdot 10^{-11}$ m is the geometrical emittance and $\beta = 10$ m is the average betatron function used at $\lambda = 1.5$ nm. Since the effect of the energy spread $\sigma_E$ over $\Delta v_z$ is $\Delta v_z/v \sim \gamma^{-2} \sigma_E/E_0 \sim 5 \cdot 10^{-12}$, we conclude that the betatron motion should not constitute a serious problem.
Also note that if the focusing system is not varied with the energy, an energy increase results into a scaling of $\epsilon \sim \gamma^{-1}$ and $\beta \sim \gamma^{-1}$, and there is no energy dependence on the effects of the betatron motion. Also note that the baseline undulator length is sufficient for operation at even higher betatron function in the order of $15 - 20$ m  and in the drift section we are free to use, if needed higher betatron functions. The choice of $\beta = 10$m at $1.5$ nm, equal to the betatron function in the baseline is discussed here for simplicity only, since it is advantageous, for feasibility study purposes to assume that there is no significant difference in the betatron function along the $20$ m-long drift section and in the SHAB undulator focusing system. From this viewpoint, the argument of the suppression factor due to the betatron motion effects goes with the square of the velocity spread yielding a dependence $\sim \beta^{-2}$ allowing for flexibility in the choice of $\beta$.

After the straight section, electron beam and radiation pass through horizontal and vertical slits, suppressing the linearly-polarized soft X-ray radiation from the LCLS baseline undulator. Since the slits are positioned 40 m downstream of the planar undulator, the radiation pulse has a ten times larger spot size compared with the electron bunch transverse size, and the background radiation power can therefore be diminished of two orders of magnitude, Fig. \ref{lclspc3}. The slits can be made of Beryllium foils, for a total thickness of $150~\mu$m. The transmission for a $150~\mu$m-thick Beryllium foil is shown in Fig. \ref{transmB} in the energy range between $600$ eV and $1600$ eV. Such a foil will block the radiation, but will let the electrons go through \cite{EMMA}, Fig. \ref{lclspc6}.

Thus, the advantage of the spoiling scheme is that radiation is attenuated of $20$ dB, but while the halo of the electron bunch is allowed to propagate through the setup up to the beam dump without electron losses. In order to understand the effects of the foils on the electrons, we need to address multiple Coulomb scattering in the foils. For estimations, we use the following formula used to calculate the rms electron beam angular divergence for the same electron beam:

\begin{eqnarray}
\sigma_\theta = \sqrt{\langle\theta^2\rangle} = [21/P(MeV/c)] \sqrt{t/X0}~,
\label{mulsc}
\end{eqnarray}
where $P$ is the electron momentum in MeV/c, $t$ is thickness of foil, and $X0$ is the radiation length of the material. For Beryllium, $X0 = 35$ cm.   For a thickness $t=150~\mu$m, this amounts to $\sigma_\theta = 0.1$ mrad. Assuming an electron beam size on the foil of $15~\mu$m rms, which can be obtained by proper tuning of the focusing system, we obtain a spoiled emittance $\epsilon \simeq 1.5$ nm, corresponding to a normalized emittance $\epsilon_n = \gamma \epsilon \simeq 13~\mu$m. Such normalized emittance is well within the acceptance of the beamline optics. It should be noted that the emittance estimation is a conservative one, since only the electron beam halo, and not the main part of the beam, passes through the foils. The energy perturbation due radiation losses through the foils can be easily estimated as $\Delta \gamma/\gamma \simeq t/X0 = 0.05 \%$. We can also estimate influence of ionization losses. The ratio of losses to ionization losses is about $Z\cdot E(MeV)/800$, where $Z$ is the atomic number and $E$ is energy of the electron beam. For Beryllium, $Z = 4$ and we have ratio of order $20$, meaning that we can neglect ionization losses.

Finally, it should be noted that in the particular case of the LCLS baseline, we have an additional suppression of a few times for the linearly polarized radiation background, which we did not actually accounted for in our calculations for the degree of polarization, due to the fact that the first LCLS undulator operates in the linear regime.

\section{\label{sims} FEL simulations}

Following the previous Section considering the method in detail and discussing estimations, here we present more detailed FEL simulations with the help of the FEL code GENESIS 1.3 \cite{GENE} running on a parallel machine. We present a statistical analysis consisting of $100$ runs. Parameters used in the simulations for the low-charge mode of operation are presented in Table \ref{tt1}. The choice of the low-charge mode of operation is motivated by simplicity.

\begin{table}
\caption{Parameters for the low-charge mode of operation at LCLS used in
this paper.}

\begin{small}\begin{tabular}{ l c c}
\hline & ~ Units &  ~ \\ \hline
Undulator period      & mm                  & 30     \\
K parameter (rms)     & -                   & 2.466  \\
Wavelength            & nm                  & 0.15   \\
Energy                & GeV                 & 13.6   \\
Charge                & nC                  & 0.02 \\
Bunch length (rms)    & $\mu$m              & 1    \\
Normalized emittance  & mm~mrad             & 0.4    \\
Energy spread         & MeV                 & 1.5   \\
\hline
\end{tabular}\end{small}
\label{tt1}
\end{table}

\begin{figure}
\includegraphics[width=1.0\textwidth]{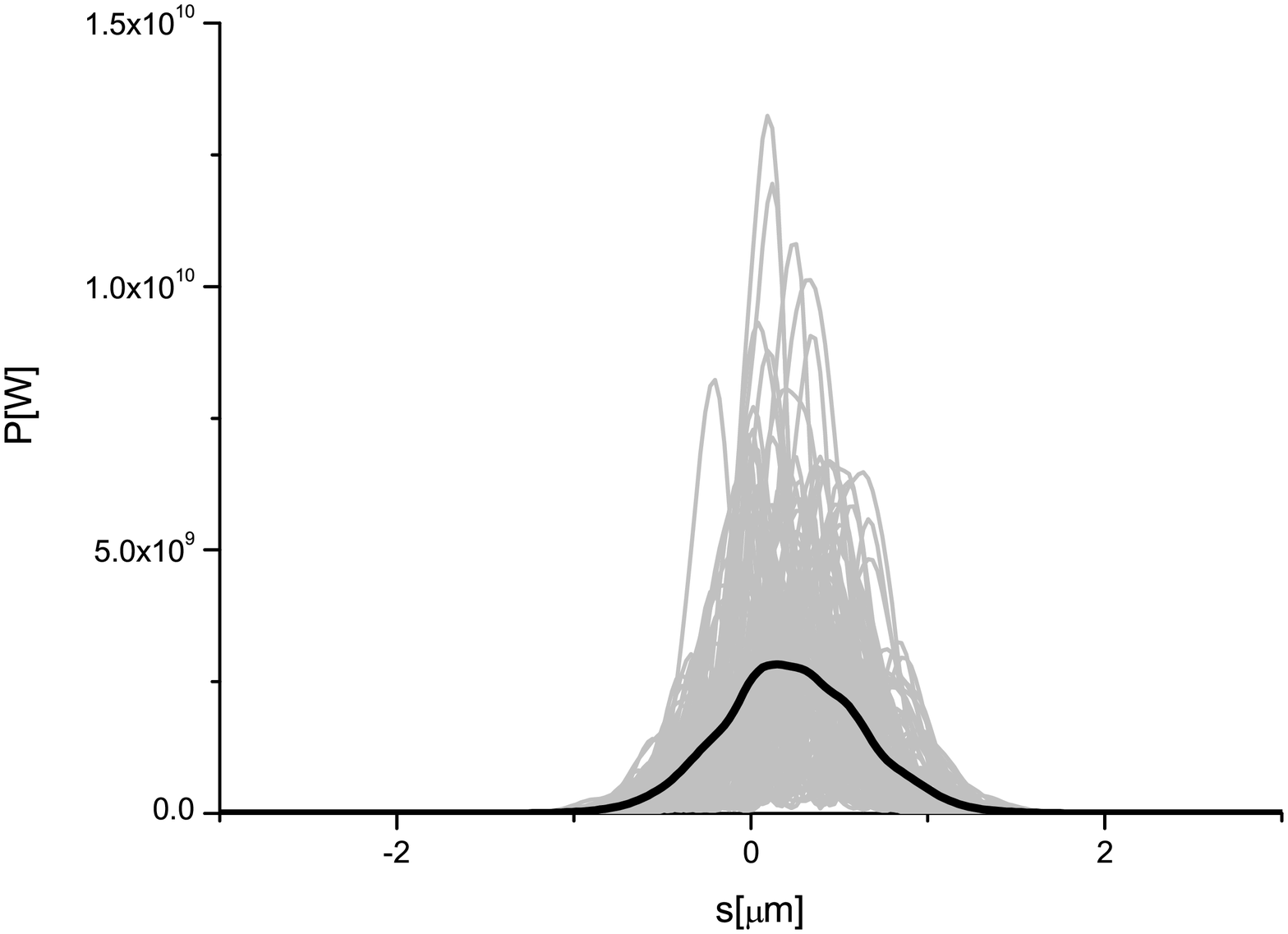}
\caption{Power distribution after the first SASE undulator (5 cells). Grey lines refer to single shot realizations, the black line refers to the average over a hundred realizations. } \label{powbuncher}
\end{figure}

\begin{figure}
\includegraphics[width=1.0\textwidth]{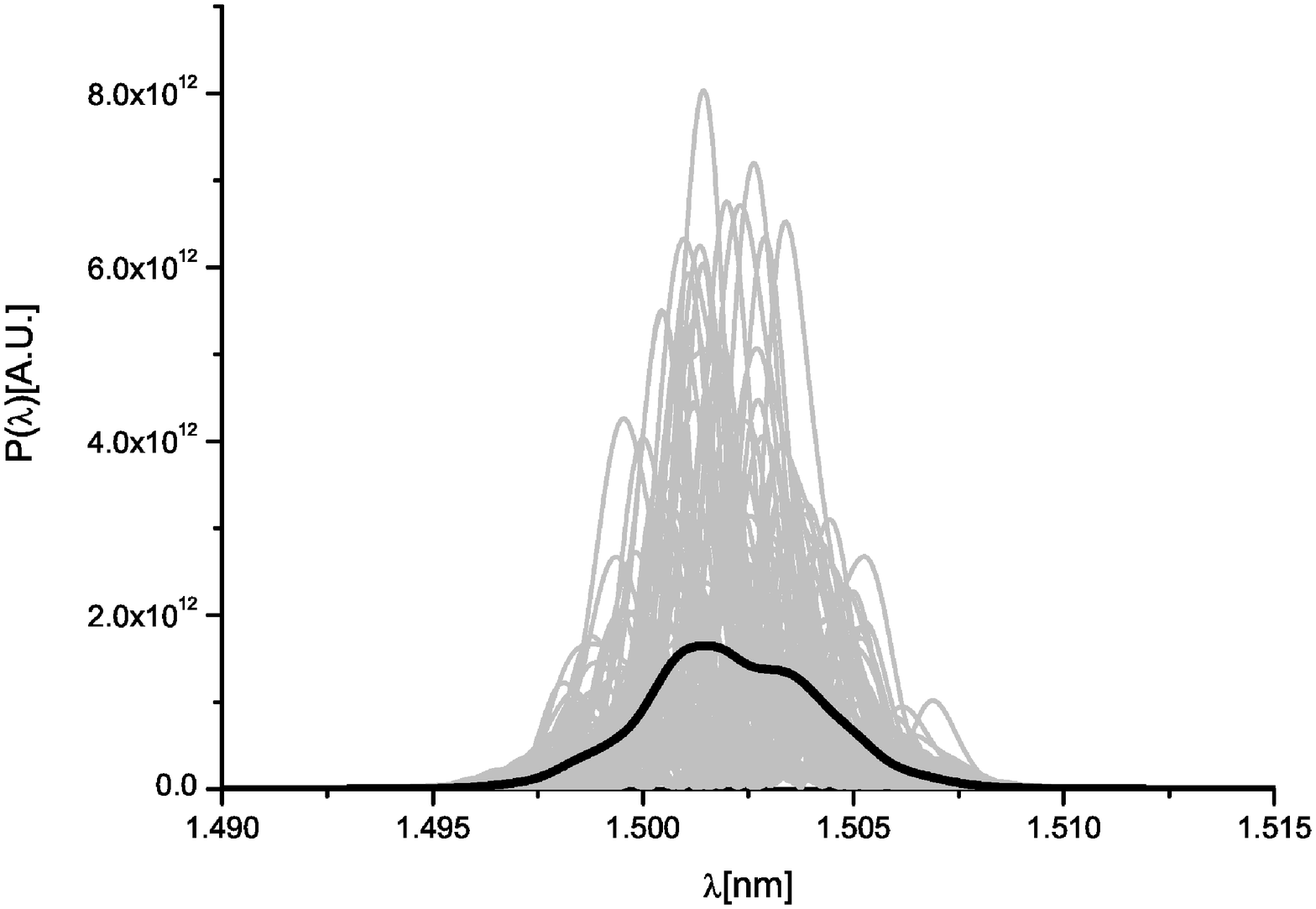}
\caption{Spectrum after the first SASE undulator (5 cells). Grey lines refer to single shot realizations, the black line refers to the average over a hundred realizations.} \label{spbuncher}
\end{figure}
\begin{figure}
\includegraphics[width=1.0\textwidth]{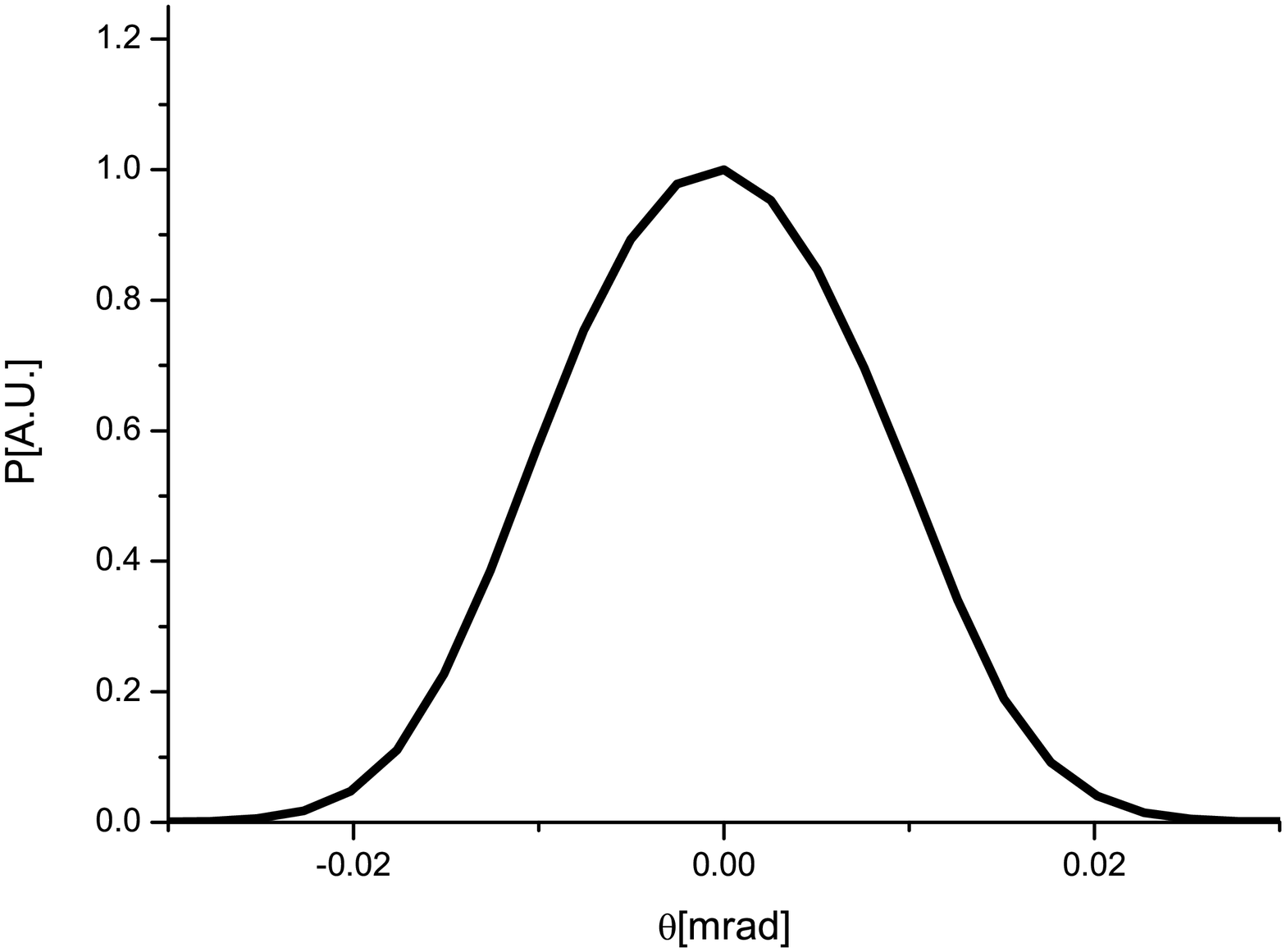}
\caption{Angular X-ray radiation pulse energy distribution after the first SASE undulator (5 cells).} \label{Angdis}
\end{figure}

As discussed in the previous section, the baseline LCLS undulator should work in the linear regime. An optimum is found when only the last $5$ cells upstream of the SHAB are used. In other words we assume that first 23 baseline undulator modules are detuned. The power and spectrum after the baseline undulator are shown in Fig. \ref{powbuncher} and Fig. \ref{spbuncher}. Fig. \ref{Angdis} shows the X-ray radiation pulse energy distribution in the far zone, which is about $20 ~\mu$rad FWHM wide.

\begin{figure}
\includegraphics[width=1.0\textwidth]{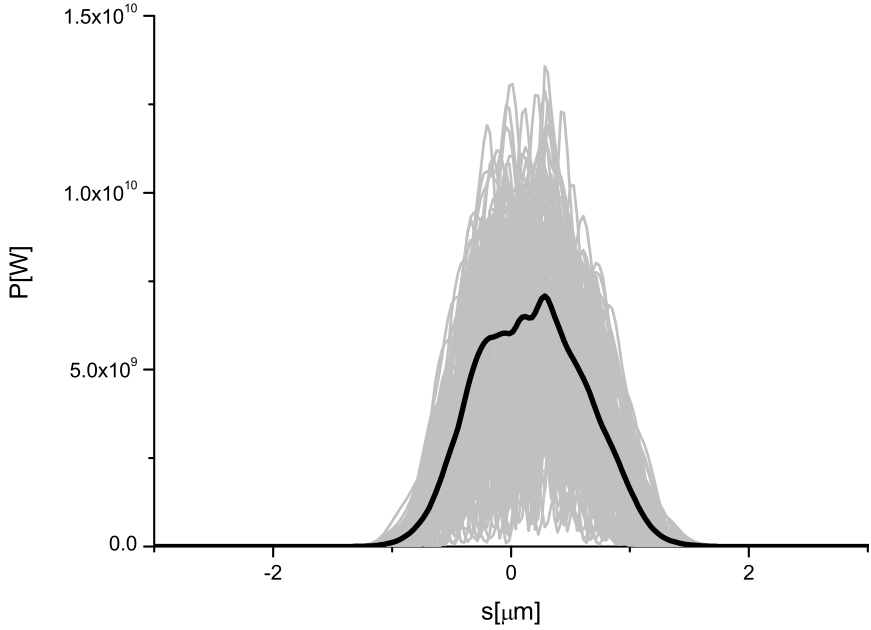}
\caption{Power after the 5 m-long APPLE type undulator. Grey lines refer to single shot realizations, the black line refers to the average over a hundred realizations.} \label{powapple}
\end{figure}

\begin{figure}
\includegraphics[width=1.0\textwidth]{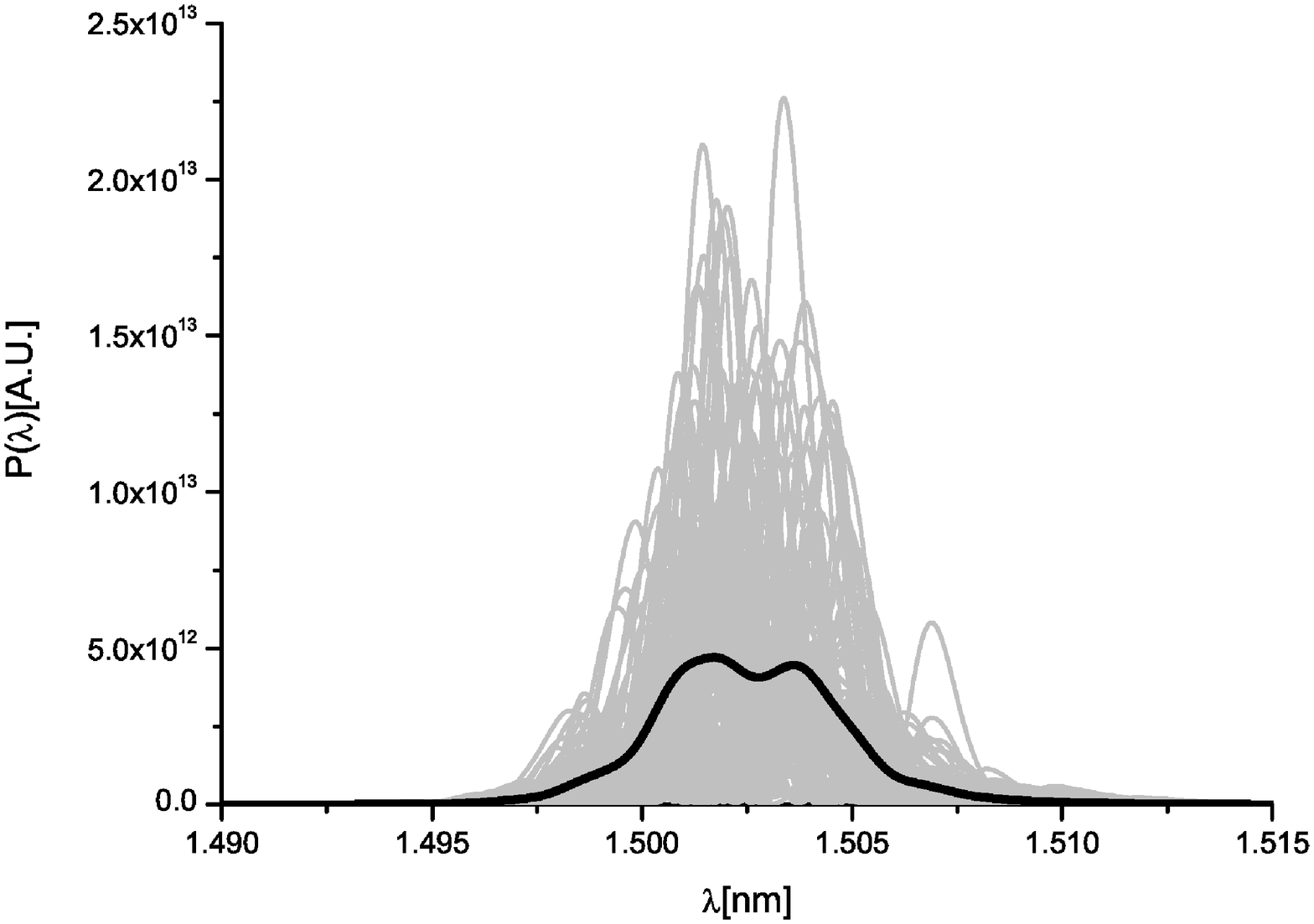}
\caption{Spectrum after the 5 m-long APPLE type undulator. Grey lines refer to single shot realizations, the black line refers to the average over a hundred realizations.} \label{spapple}
\end{figure}

\begin{figure}
\includegraphics[width=1.0\textwidth]{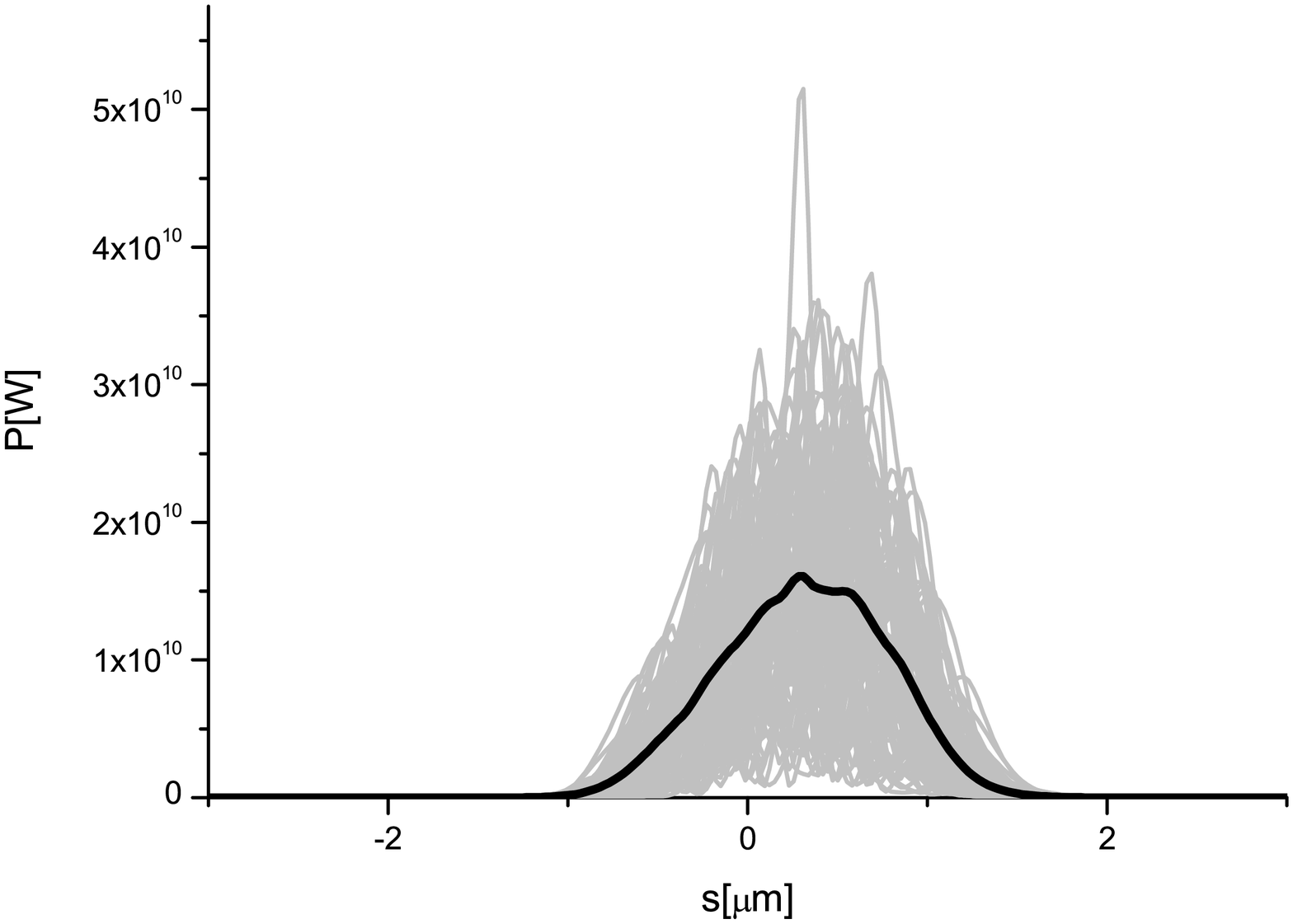}
\caption{SASE power from the LCLS baseline at saturation (6 modules). Grey lines refer to single shot realizations, the black line refers to the average over a hundred realizations.} \label{Power6}
\end{figure}

\begin{figure}
\includegraphics[width=1.0\textwidth]{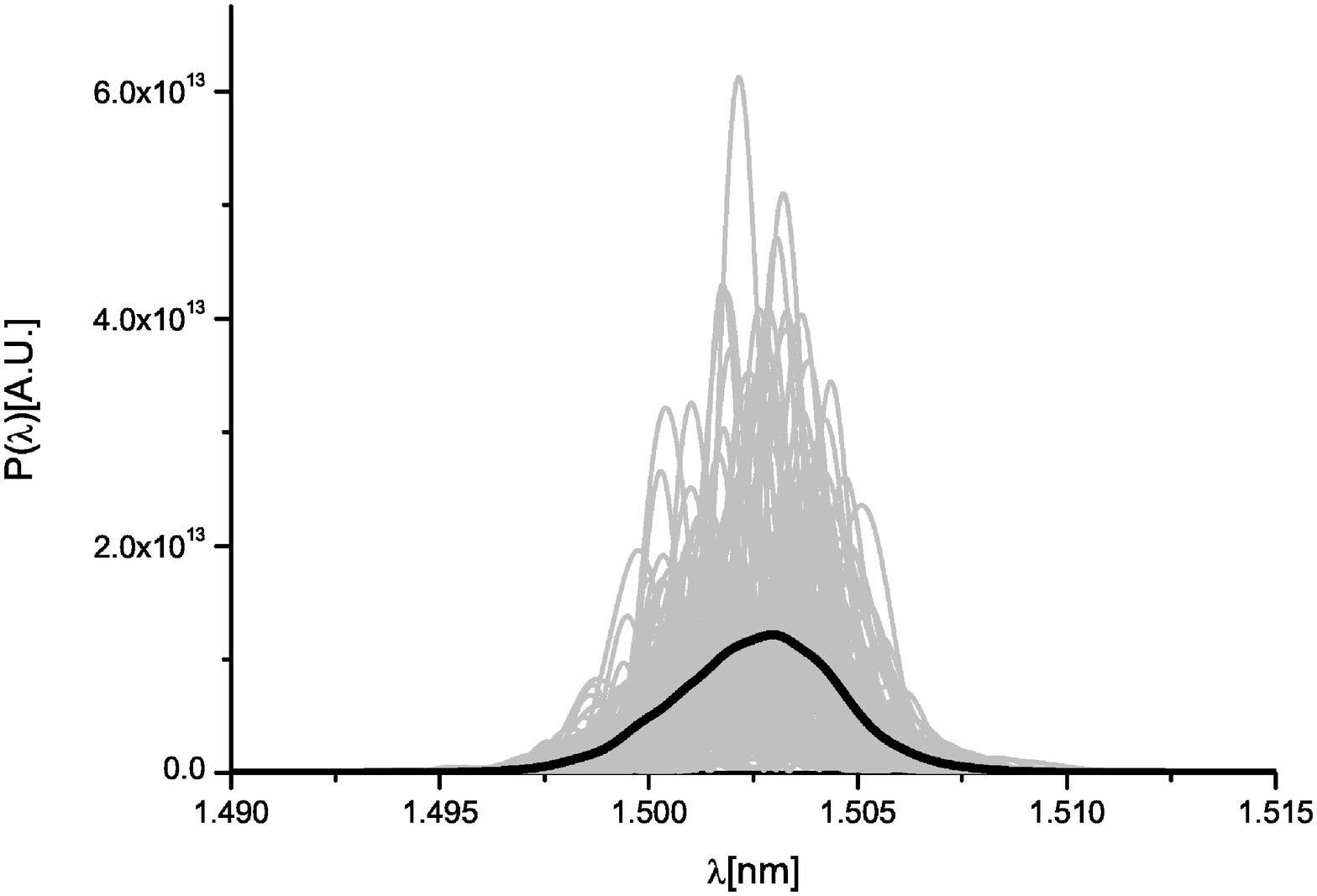}
\caption{SASE spectrum from the LCLS baseline at saturation (6 modules). Grey lines refer to single shot realizations, the black line refers to the average over a hundred realizations.} \label{Spectrum6}
\end{figure}

The particle file produced by Genesis at the exit of baseline is subsequently transformed assuming  a dispersive element with $R_{56} \simeq L/\gamma^2 \simeq 560$ nm, and used as an input for further simulations through the $5$ m-long APPLE type undulator. The average betatron function is assumed to be $\beta = 10$ m. From a practical viewpoint, this means that we assume that the same focusing system in the SHAB section is continued through the following $20$ m-long straight section up to the APPLE, which is installed after the last quadrupole. Such assumption can obviously be relaxed, and it is considered here for simplicity reasons only. Also, the influence of the betatron motion on the microbunching is only estimated in the previous Section, but is not explicitly accounted for in simulations. However, from previous estimations we expect that such influence would be even smaller than the influence due to the finite energy spread of the beam. Moreover, as note before, the betatron function can be increased with respect to our choice, up to a value of $30$ m, decreasing the influence of the betatron motion on the microbunching. In this case, a small value of $\beta$ should be organized locally at the slit position. The power and spectrum after the APPLE undulator is shown in Fig. \ref{powapple} and Fig. \ref{spapple}. It can be seen from these figure that our scheme is capable of providing $10$-GW level, $99\%$ circularly polarized radiation pulses at the fundamental harmonic of $1.5$ nm. It is instructive to compare the output of our device with the power and spectrum, linearly polarized, from the baseline LCLS undulator at saturation, Fig. \ref{Power6} and Fig. \ref{Spectrum6} respectively, which is comparable with the output from our setup.

%

%


\section{Conclusions}

Enhancing the capabilities of the operating LCLS baseline is a challenging problem, subject to many constraints including low cost, little available time to perform changes and guarantee of a safe return to the baseline mode of operation. In this paper we propose a method of polarization control which offers simplicity and flexibility, and can be added to the existing LCLS X-ray FEL without significant cost or design changes. The setup can be installed in a little time and is not expensive. Implementation of the proposed technique downstream of the baseline undulator will not perturb the baseline mode of operation of the LCLS undulator. Moreover, at present, detailed experience is available in synchrotron radiation laboratories concerning the manufacturing of proposed APPLE II undulator radiator. An improved technology in the APPLE II undulator design has been enabled in recent years in well-experienced laboratories (consider, e.g. the development of the APPLE undulators for PETRA III) and  this insertion device has meanwhile become commercially available, with a manufacturing time which can be estimated in two years. Altogether, we offer an alternative scheme to currently available methods for polarization control which promises excellent, cost-effective, and risk-free results.

\section{Acknowledgements}

We are grateful to Massimo Altarelli, Reinhard Brinkmann, Serguei
Molodtsov and Edgar Weckert for their support and their interest
during the compilation of this work.

\end{document}